\documentstyle[color,amstex,amssymb,12pt,psfig,epsf]{article}
\author{Antonio Coniglio, Annalisa Fierro, Mario Nicodemi\\
\small Dipartimento di Fisica, Universit\'{a} ``Federico II'', \\
\small INFM and INFN Napoli, Via Cintia, 80126 Napoli, Italy}
\title{Stationary Probability Distribution in Granular Media}
\date{3/2/2003}

\newcommand{\e}{\mbox{e}}
\newcommand{\lan}{\langle}
\newcommand{\ran}{\rangle}

\begin{document}
\maketitle
\begin{abstract} We discuss recent developments in the formulation of 
a Statistical Mechanics approach to non thermal systems, such as granular 
media. We review a few important numerical results on the assessment of 
Edwards' theory and, in particular, we apply these ideas to study 
a mean field model of a hard sphere binary mixture under gravity, 
which can be fully analytically 
investigated. As a consequence, we derive the rich phase diagram and 
predict the features of segregation patterns of the mixture. 
\end{abstract}

\section{Introduction}
The issue we discuss here, which recently raised considerable
interest \cite{Edwards,fnc,mimmo,kurchan,Dean,brey}, 
is the possibility to describe non thermal systems, 
such as granular media, by using concepts from Statistical Mechanics, 
as proposed by Edwards in 1989 \cite{Edwards}. 

\begin{figure}[ht]
\centerline{\psfig{figure=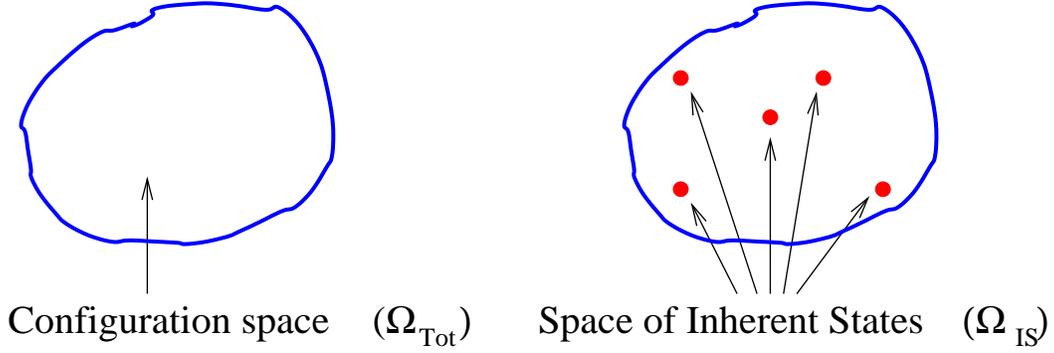,width=14cm,angle=0}}
%\vspace{-1cm}
\caption{A schematic picture of the system whole configuration space, 
$\Omega_{Tot}$, and the subspace of {\em Inherent States}, $\Omega_{IS}$, 
i.e., mechanically stable states.} 
\label{space}
\end{figure}

Granular media  \cite{JNB} are dissipative systems made of macroscopic grains, 
such as powders or sands, where gravity effects are usually much stronger 
than thermal agitation, which is thus neglected (i.e., $T_{bath}=0$). 
As much as thermal systems,
their macroscopic properties are characterized by a few control parameters 
and their macrostates correspond to a huge number of microstates.  
In granular media at rest, these microstates are 
mechanically stable, ``frozen'', arrangements of grains. 
%corresponding to local minima (or saddles) of the energy. 
They have been called {\bf ``inherent states''} \cite{fnc}, 
(see Fig.\ref{space}), 
because of their connections to ``inherent structures'' \cite{Stillinger} 
of supercooled liquids quenched at zero temperature, i.e., 
local minima of the potential energy in configuration space (which were 
introduced for their important role in glassy properties). 

In thermal systems the space of microstates is explored by the presence 
of a finite $T_{bath}$. In granular media at rest $T_{bath}=0$, but the 
effects of a finite thermal bath ($T_{bath}>0$) can be obtained by an external 
drive (like a ``shake'') of amplitude $\Gamma>0$.

The above considerations outline the possibility to develop a 
Statistical Mechanics approach to describe granular media. 
In the present paper, this approach is introduced in Sect. \ref{sect1} 
and its check, via Monte Carlo simulations in the schematic framework 
of lattice models for granular media, is reviewed in Sect. \ref{sect2}. 
Sect. \ref{sect3} is devoted to the analytical investigation of a 
mean field model of a binary hard sphere mixture under gravity 
for granular media, treated with the above approach. 
Finally, mean field predictions about the system phase diagram and its 
segregation phenomena are discussed in Sect. \ref{sect4}. 

\begin{figure}[ht]
\vspace{1cm}
\centerline{\psfig{figure=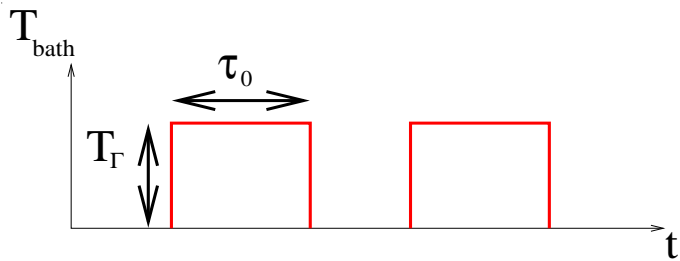,width=9cm,angle=0}}
\caption{Our lattice grain models are subject to a Monte Carlo dynamics 
made of ``tap'' sequences. A ``tap'' is a period of 
time, of length $\tau_0$ (the tap duration), where they evolve 
at a finite value of the bath temperature, $T_{\Gamma}$ (the tap amplitude); 
afterwards grains are frozen in one of their inherent states. } 
\label{fig_tap_mc}
\end{figure}

\section{Probability distribution}
\label{sect1}
Edwards \cite{Edwards} proposed a method to individuate 
the probability, $P_r$, to find the system in its inherent state $r$, 
under the assumption that these mechanically
stable states have the same a priori probability to occur. 
The knowledge of $P_r$ has the conceptual advantage to substitute {\em time} 
with {\em ensemble averages}, and thus allowing the description of the 
system properties by use of few basic theoretical concepts, as in 
thermodynamics. A possible approach to find $P_r$ is as follows \cite{fnc}. 
$P_r$ is obtained as the maximum of the entropy, 
\begin{equation}S=-\sum_r P_r\ln P_r\end{equation}
with the macroscopic constraint that 
the system energy, $E = \sum_r P_r E_r$, is given.
This assumption leads to the Gibbs result:
\begin{eqnarray}
P_r\propto e^{-\beta_{conf} E_r}
%\nonumber
\label{pr}
\end{eqnarray}
where $\beta_{conf}$ is a {\em Lagrange multiplier}, called {\em inverse 
configurational temperature}, enforcing the above constraint 
on the energy: 
\begin{equation}
\beta_{conf}= \frac{\partial S_{conf}} {\partial E}
\ \ \ \ \ \ \ \ \  S_{conf} = \ln \Omega_{IS} (E)
\end{equation}
Here, $\Omega_{IS}(E)$ is the number of inherent states with energy $E$.
Thus, summarizing, the system at rest has 
%\begin{center}
$T_{bath} = 0$ %\hspace{0.5cm} 
and %\hspace{0.5cm}
$T_{conf} =  \beta_{conf}^{-1} \neq 0$.
%\end{center}

\begin{figure}[ht]
\centerline{\psfig{figure=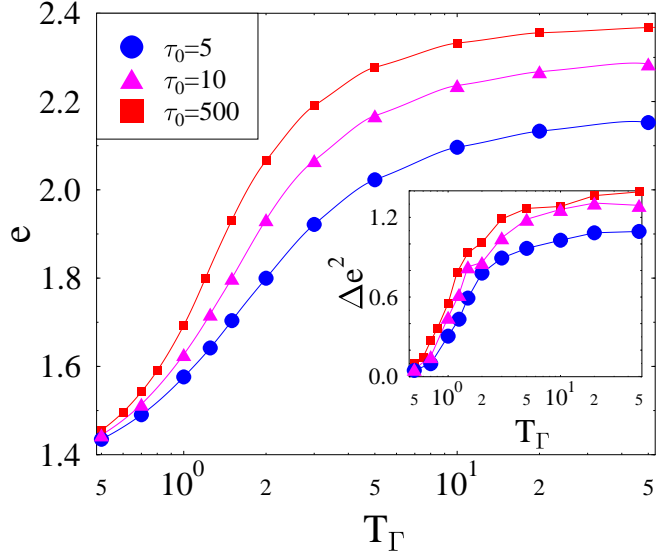,width=8cm,angle=-90}}
\vspace{-2cm}
\caption{The time average of the energy, $e$, and ({\bf inset}) its
fluctuations, $\Delta e^2$, recorded at stationarity during
a tap dynamics, as a function of the tap amplitude, $T_\Gamma$, 
in the 3D lattice monodisperse hard sphere model. 
Different curves correspond to sequences of tap with
different values of the duration of each single tap, $\tau_0$.} 
\label{fig_e_de_tgamma}
\end{figure}

These basic considerations, to be validated by experiments or simulations, 
settle a theoretical Statistical Mechanics framework to describe granular 
media. Consider, for definiteness, a system of monodisperse hard spheres 
of mass $m$. In the system whole configuration space $\Omega_{Tot}$, 
we can write Edwards' generalized partition function as: 
\begin{eqnarray}
Z=\sum_{r\in\Omega_{Tot}} \e^{-{\cal H}_{HC}-\beta mgH}
\cdot \Pi_r(\gamma) %\nonumber
\label{Z_thedw}
\end{eqnarray}
where ${\cal H}_{HC}$ is the hard core interaction between grains, $mgH$ is 
the gravity contribution to the energy ($H$ is particles height), 
and the factor $\Pi_r(\gamma)$ is a projector on the inherent states space 
$\Omega_{IS}$ (see Fig.\ref{space}): 
if $r\in\Omega_{IS}$ then $\Pi_r(\gamma)=1$ else 
$\Pi_r(\gamma)=1-\gamma$. 
Usual Statistical Mechanics, where $\beta^{-1}$ is identified with 
$T_{bath}$, is recovered for $\gamma=0$; Edwards' ``Inherent States'' 
Statistical Mechanics is obtained for $\gamma\rightarrow 1$, and 
by definition $\beta^{-1}$ is called $T_{conf}$ \cite{Edwards,fnc}.

\begin{figure}[ht]
\centerline{\psfig{figure=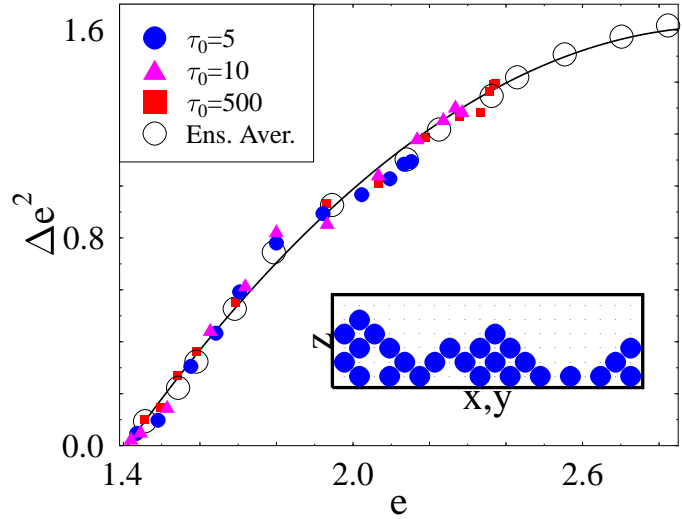,width=8cm,angle=-90}}
\vspace{-2cm}
\caption{The time averages of energy fluctuations, $\Delta e^2$, 
when plotted as function of the time average of energy, $e$, 
collapse on a single master function for all the different
values of tap amplitude and duration, $T_\Gamma$ and $\tau_0$, shown
in Fig.\ref{fig_e_de_tgamma}. Ensemble averages independently calculated 
over the distribution eq.(\ref{pr}), $\lan \Delta e^2 \ran$, 
go on the same curve when plotted as a function of $\lan e \ran$
(the black empty circles). Time and Edwards ensemble averages coincide 
to a very good approximation. 
Data refer to a 3D monodisperse hard-sphere model under gravity, 
sketched in the {\bf inset}.} 
\label{fig_e_de}
\end{figure}

\section{Lattice hard sphere models for granular media}
\label{sect2}
The above scenario has been checked \cite{fnc} by Monte Carlo simulations, 
in a schematic model for granular media: 
a system of monodisperse hard-spheres (with diameter $\sqrt{2}a_0$) under
gravity, where the centers of mass of grains are constrained to move 
on the sites of a cubic lattice (of spacing $a_0=1$, 
see inset Fig.\ref{fig_e_de}). 
Grains are subject to a dynamics made of a sequence of ``taps'' \cite{NCH} 
(see Fig.\ref{fig_tap_mc}): a single ``tap'' is a period of time, 
of length $\tau_0$ (the tap duration), where they evolve 
at a finite value of the bath temperature, $T_{\Gamma}$ (the tap amplitude); 
afterwards grains are frozen in one of their inherent states. 

We showed  that under such a tap dynamics the systems reaches 
a stationary state where time averages can be replaced, 
with good agreement, by ensemble averages over the above 
generalized Gibbs distribution eq.(\ref{pr}). 

\begin{figure}[ht]
\centerline{\psfig{figure=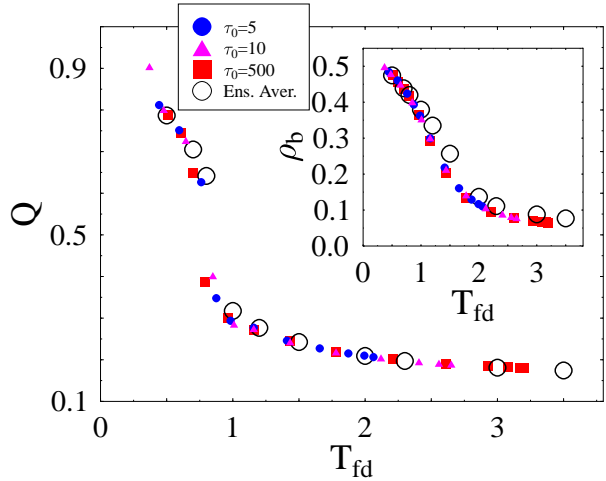,width=8cm,angle=-90}}
%\vspace{-1cm}
\caption{The density self-overlap function, $Q$, and ({\bf upper inset})
the system density on the bottom layer, $\rho_b$, (previously recorded 
during tap dynamics with different durations $\tau_0$ and amplitude 
$T_{\Gamma}$) are 
plotted as function of $T_{fd}$ (in units $mga_0$). These time averages are 
compared with the ensemble averages over the distribution eq.(\ref{pr})
(the black empty circles), plotted as function of $T_{conf}$ (in units
$mga_0$). There is a very good agreement between the
independently calculated time averages over the tap dynamics and the
statistical mechanics ensemble averages \'a la Edwards.
Data refer to a 3D monodisperse hard-sphere model under gravity.} 
\label{fig_mon_hs}
\end{figure}

Time averages over the tap dynamics at stationarity of the energy, $e$, 
and energy fluctuations, $\Delta e^2$, as a functions of $T_{\Gamma}$, are 
plotted in Fig.\ref{fig_e_de_tgamma}, where different curves correspond to 
tap sequences of different tap duration $\tau_0$. Apparently, $e$ and 
$\Delta e^2$ are functions of both $T_{\Gamma}$ and $\tau_0$, thus, for 
instance, $T_{\Gamma}$ does not result to be 
a good parameter to uniquely individuate the stationary status 
of the system. Anyway, we can see that a single thermodynamic parameter 
is necessary to 
characterize the macroscopic status. This is apparent from Fig.\ref{fig_e_de}
(see also Fig. \ref{fig_mon_hs}), 
where we plot $\Delta e^2$ as a function of $e$: all the data from 
different tap dynamics of Fig.\ref{fig_e_de_tgamma} 
collapse on a single master curve. 
We independently calculated the ensemble averages over distribution 
(\ref{pr}) of $e$ and $\Delta e^2$: interestingly these collapse on the 
same curve drawn by data from tap dynamics (see Fig.\ref{fig_e_de}). 

Alternatively, an intensive ``thermodynamic'' parameter 
can be defined from the data collected along the tap dynamics at stationarity 
through the static fluctuation dissipation 
relation, which leads to individuate a ``temperature'', $T_{fd}$ \cite{fnc}. 
We numerically showed that $T_{fd}$ coincides with Edwards' 
``configurational temperature'', $T_{conf}$, defined above. 
These results are briefly summarized in Fig. \ref{fig_mon_hs}. 
There we show that averages of macroscopic quantities calculated 
along different tap dynamics %(i.e., tap sequences with different 
%tap duration, $\tau_0$, or amplitude, $T_{\Gamma}$) 
collapse on a 
single master curve when plotted as a function of the ``thermodynamic'' 
parameter $T_{fd}$. In Fig. \ref{fig_mon_hs}, we also show that 
ensemble averages, independently calculated by eq.(\ref{pr}), collapse 
on the very same master curve when plotted as a function of $T_{conf}$. 

\begin{figure}[ht]
\centerline{\psfig{figure=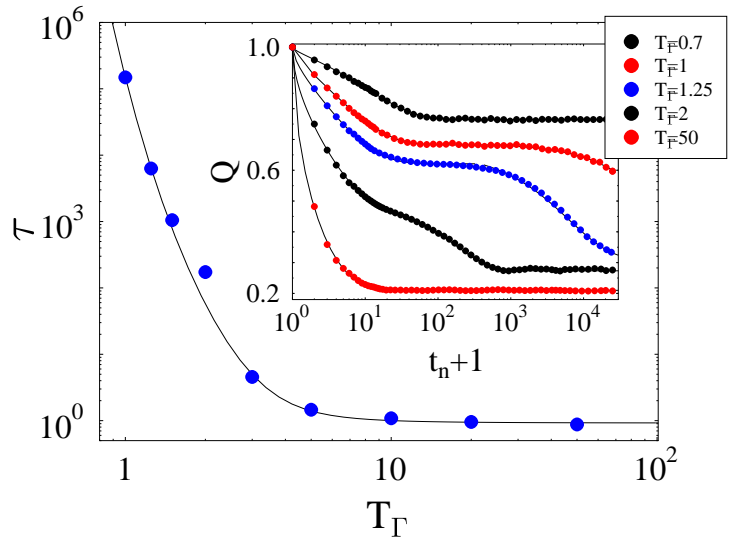,width=8cm,angle=-90}}
\vspace{-2cm}
\caption{{\bf Inset} The density self-overlap, $Q$, as a function of the 
number of taps, $t_n$, for several $T_\Gamma$. 
{\bf Main Frame} The characteristic relaxation time, $\tau$, defined 
by a stretched exponential fit of the long time dynamics of $Q(t_n)$. 
At low $T_\Gamma$, 
$\tau$ appear to diverge \'a la Arrhenius or Vogel-Tamman-Fulcher (the 
superimposed fit).} 
\label{fig_q_t}
\end{figure}

The tap ``dynamics'' is interesting in itself \cite{NCH,mimmo,cn_g}. 
For instance, density relaxation is well fitted by stretched exponential 
and by logarithmic laws respectively at high and low $T_{\Gamma}$ 
\cite{fnc,NCH}, a fact also experimentally found \cite{Knight}. 
In particular, one observes 
that by lowering $T_{\Gamma}$ the systems gets ``jammed'', in the sense that 
its characteristic relaxation time, $\tau$, enormously grows. This 
is shown in Fig.\ref{fig_q_t}. 
The presence of strong ``aging'' effects in models for granular media 
was indeed discovered a few years 
ago \cite{mimmo,cn_g}. The growth of $\tau$ with $T_{\Gamma}$ is 
well approximated by an Arrhenius or Vogel-Tamman-Fulcher law \cite{fnc,NCH}, 
resembling the slowing down of glass formers close to the glass transition, 
a result recently experimentally confirmed \cite{danna}. 

\begin{figure}[ht]
\vspace{1cm}
\centerline{\psfig{figure=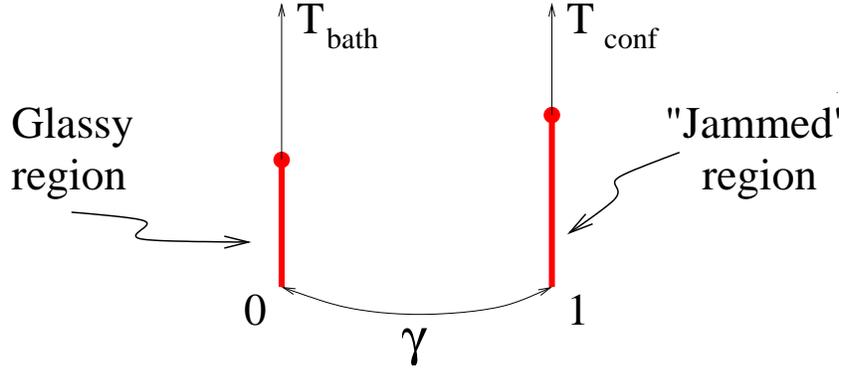,width=11cm,angle=0}}
%\vspace{-0.5cm}
\caption{A schematic picture of a unifying phase diagram for the 
``glassy region'' of ordinary hard spheres under gravity and 
``jammed region'' of granular media which emerges from the 
Statistical Mechanics approach to powders. 
%Also the location of the ``Brazil Nut Effect'' (BNE) 
%and ``Reverse Brazil Nut Effect'' (RBNE) regions is shown.
} 
\label{unif_phdg}
\end{figure}

Eq.(\ref{Z_thedw}) gives a general framework where the properties of thermal 
and non-thermal systems, such as their phase diagram, can be 
discussed. For instance, 
a relation, as the one sketched in Fig.\ref{unif_phdg}, is now found 
to exist between glassy properties of hard spheres under gravity in usual 
statistical mechanics ($\gamma=0$ limit) and the ``jamming'' properties 
of granular media ($\gamma=1$ limit);
the control parameters being the number of grains and, respectively 
the temperature, $T_{bath}$, and the configurational temperature, $T_{conf}$. 
This is consistent with the theoretical speculation by Liu and Nagel \cite{LN} 
about the existence of a generalized phase diagram for the ``jammed'' region 
of granular media and ``glassy'' region of glass formers. 

Also in glasses, following for example the inherent structure approach
\cite{Stillinger}, one can
define  a configurational entropy associated to the number of
inherent structures corresponding to a fixed energy, $E$, and consequently the
configurational temperature. When the system is frozen at zero temperature
in one of its inherent states it does not evolve anymore. A 
way to visit the inherent structures is by using a tap dynamics 
(i.e., a procedure similar to that used above for granular materials).
Thus, more generally, the framework of eq.(\ref{Z_thedw}) opens the way 
to a Statistical Mechanics of ``frozen system''. 

Finally, it is important to recall that we have shown \cite{fnc} 
that more than one Lagrange multiplier may be necessary, in general, 
to assign the macroscopic system status. For instance, in the case of a 
binary mixture of hard spheres on a lattice, at least 
{\em two} configurational temperatures must be introduced. In this case, 
an extension of Edwards' approach is required, where $P_r$ is given by 
the maximum of the entropy with separate constraints on the gravitational 
energies of the two species of grains:
$E_1 = m_1g\sum_r P_r H_{1r}$ and $E_2 = m_2g\sum_r P_r H_{2r}$
($H_1$ and $H_2$ are the heights of species 1 and 2).
This gives {\em two} Lagrange multipliers:
\begin{eqnarray}
\beta_1 = \frac{\partial \ln \Omega_{IS} (E_1,E_2) }{\partial E_1}\quad
\beta_2 = \frac{\partial \ln \Omega_{IS} (E_1,E_2) }{\partial E_2} \nonumber
\end{eqnarray}
\noindent
where $\Omega_{IS}(E_1,E_2)$ is the number of inherent states with $E_1$,$E_2$.
%In the whole configuration space $\Omega_{Tot}$, 
Edwards' generalized partition function must now be written as:
\begin{eqnarray}
Z=\sum_{r\in\Omega_{Tot}} \e^{-{\cal H}_{HC}-\beta_1m_1gH_1-\beta_2m_2gH_2}
\cdot \Pi_r(\gamma) \nonumber
\end{eqnarray}
%The factor $\Pi_r(\gamma)$ is a projector on the inherent states space 
%$\Omega_{IS}$ (see Fig.\ref{space}): 
%if $r\in\Omega_{IS}$ then $\Pi_r(\gamma)=1$ else $\Pi_r(\gamma)=1-\gamma$. 

To illustrate the theoretical developments allowed by Edwards formulation, 
we now discuss a mean field model of the above binary mixture. 
We can thus predict the structure of its phase diagram and describe an 
ubiquitous and intriguing phenomenon, called ``size segregation'', observed 
in vibrated granular mixtures where in presence 
of shaking the system is not randomized, but its components tend to separate. 

\section{A Mean Field model}
\label{sect3}
We consider a lattice of $H$ layers, occupied by a hard spheres system made of 
two species, 1 (small) and 2 (large), with grain diameters $a_0$ 
and $\sqrt{2} a_0$ (the lattice spacing is $a_0=1$). 
The hard core potential ${\cal H}_{HC}$ is such that two nearest neighbor 
particles cannot overlap. This implies that only pairs of small particles 
can be nearest neighbors on a layer.

\begin{figure}[ht]
%\vspace{-2.5cm}
\centerline{%\hspace{-5cm}
\psfig{figure=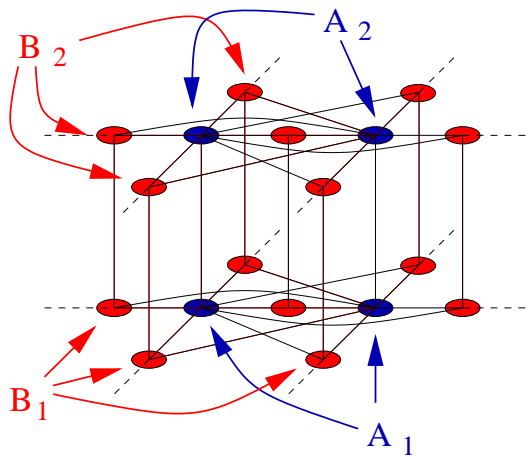,width=7cm,angle=0}}
%\vspace{-1.5cm}
\caption{A schematic picture of the lattice considered here. We have 
$H$ horizontal layers, each divided in two reciprocally fully connected 
sublattices $A$ and $B$. Grains are located on the lattice sites and, 
in order to be ``stable'', must rest on their homologous underneath grains.}
\label{mf_lat}
\end{figure}

Each layer is dealt with the standard mean field technique, i.e., it is 
subdivided in two sublattices $A$ and $B$, where each site in sublattice $A$ 
(resp. $B$) interacts with all sites in $B$ (resp. $A$) 
and only with them (see Fig.\ref{mf_lat}). 
The Hard-Core lattice Hamiltonian for spheres on the same layer at height $z$ 
(with $z\in\{1,...,H\}$) is thus:
\begin{equation}
{\cal H}_{HC}(z) = \frac{J}{2N} \sum_{ij} f(n_i^{A_z},n_j^{B_z})
\label{H_HC}
\end{equation}
Here we introduce the notation $n_i^{A_z}$ (resp.  $n_i^{B_z}$) to individuate 
the status of site $i$ in sublattice $A$ (resp. $B$) of the layer at hight $z$:
$n_i^{A,B}\in\{0,1,2\}$ is an occupancy variable which is 0 if no grains 
are on site $i$, 1 or 2 whether a small or a large sphere is there. 
In eq.(\ref{H_HC}) the sum is on all sites of sublattices $A_z$ and $B_z$, 
$i,j\in\{1,...N\}$. 
The ``shape factor'', $f(n_i,n_j)$, indicates whether two grains can be 
nearest neighbor: 
$f(n_i,n_j)=(1-\delta_{n_i,0})(1-\delta_{n_j,0})-\delta_{n_i,1}\delta_{n_j,1}$.
%\begin{equation}
%f(n_i,n_j)=-{n_in_j\over 4}\left[3n_in_j-5(n_i+n_j)+7\right]=
%(1-\delta_{n_i,0})(1-\delta_{n_j,0})-\delta_{n_i,1}\delta_{n_j,1}
%\end{equation}
The Hard-Core limit is found for $J\rightarrow\infty$. 

To set the average density a chemical potential term is also present:
\begin{equation}
{\cal H}_{\mu}(z) = -\sum_{i,a}\left(\mu_1\delta_{n_i^a,1}+
\mu_2\delta_{n_i^a,2}\right)
\label{H_mu}
\end{equation}
Here the sum is on all sites of sublattices $A_z$ and $B_z$ (i.e, 
$i\in\{1,...N\}$ and $a\in\{A_z,B_z\}$).
Of course, also the gravity contributions are considered:
\begin{equation}
%{\cal H}_{g}(z) = g \sum_{i,a} m(n_i^a)z_i^a=
%m_1gz\sum_{i,a} \delta_{n_i^a,1}+m_2gz\sum_{i,a} \delta_{n_i^a,2}=
%{\cal H}_{g}^{(1)}(z)+{\cal H}_{g}^{(2)}(z)
{\cal H}_{g}^{(1)}(z)=m_1gz\sum_{i,a} \delta_{n_i^a,1} \ \ \ \ ; \ \ \ \ 
{\cal H}_{g}^{(2)}(z)=m_1gz\sum_{i,a} \delta_{n_i^a,2}
\label{H_g}
\end{equation}
where $z$ is the height of the layer where $n_i^a$ belong 
to, and $m_k$, $k\in\{1,2\}$, is the mass of grains of type 1 or 2. 

\subsection{Edwards projector term ${\cal H}_{Edw}$}

The above Hamiltonian is the usual one describing a set of hard spheres 
under gravity at a mean field level. The novelty we consider here is 
Edwards' idea that in granular media averages must be done by considering 
only mechanically stable states (i.e., ``Inherent States'', IS). 
In the present case of a lattice model the ``stability'' of a grain 
is given by the presence of some underlying grains which support it. 
The projector on IS has thus a simple form: it can be written as a sum of 
single particle terms to be added in the Hamiltonian. 
In fact, to decide whether a grain on layer $z$ is ``mechanically stable'' 
we must just check the status of its corresponding sites in the two layers 
below it. 
In the case of not fully connected layers also next neighbors sites at level 
$z-1$ should be considered, but this is not the case here. 
In particular, since for a matter of simplicity we just discuss a system 
made of 2 layers (i.e., $H=2$), only the status of site $i$ at $z-1$ must 
be considered to establish whether $n_i$ at $z$ is stable or not. 
The case with more than 2 layers gives very similar results to those 
presented here, but at the price of a more intricate analytical treatment.
Since our model is already very schematic, we disregards these details and 
restrict here to $H=2$. Thus, summarizing, in the present case 
the ``stability'' of grains on level $z$ is given by the following term, 
which couples adjacent layers between them, to be added to the 
Hamiltonian \cite{note_p}:
\begin{equation}
{\cal H}_{Edw}(z)=
K \sum_{i,a}(1-\delta_{n_i^{a_z},0})\delta_{n_i^{a_{z-1}},0}
\label{H_pi}
\end{equation}
%This expression defines a grain at site $i$ on level $z$ 
%as ``mechanically stable'' (i.e., ${\cal H}_{Edw}(n_i^{a_z})=0$) 
%if the homologous site $i$ at $z-1$ is filled. 

Usual Statistical Mechanics is recovered for $K=0$. 
Edwards' ``Inherent States'' 
Statistical Mechanics is obtained for $K\rightarrow\infty$ 
(i.e., $K=\ln(1-\gamma)$); the projector on Inherent States being:
$\Pi(\{n^{a_z}_i\},K)=\exp\left\{-\sum_{z=1}^H{\cal H}_{Edw}(z)\right\}$.
%The limit $K\rightarrow\infty$ must be taken after the Hard-Core limit 
%$J\rightarrow\infty$. 
The system partition function can thus be written as:
\begin{eqnarray}
Z
%&=&\sum_{\{n^{a_z}_i\}} \e^{-{\cal H}_{HC}-\beta_1{\cal H}_g^{(1)}
%-\beta_2{\cal H}_g^{(2)}-{\cal H}_{\mu}}\cdot\Pi(\{n^{a_z}_i\},K)\nonumber\\
&=& \sum_{\{n^{a_z}_i\}} \exp\left\{-\sum_{z=1}^H \left[
{\cal H}_{HC}(z)+\beta_1{\cal H}_g^{(1)}(z)+\beta_2{\cal H}_g^{(2)}(z)+
{\cal H}_{\mu}(z)+{\cal H}_{Edw}(z)\right]\right\}
\label{zedw}
\end{eqnarray}
%where we have defined ${\cal H}_{HC}=\sum_{z=1}^H {\cal H}_{HC}(z)$,
% ${\cal H}_{g}^{(1,2)}=\sum_{z=1}^H {\cal H}_{g}^{(1,2)}(z)$, 
%${\cal H}_{\mu}=\sum_{z=1}^H {\cal H}_{\mu}(z)$ and
%${\cal H}_{Edw}=\sum_{z=1}^H {\cal H}_{Edw}(z)$.

\begin{figure}[ht]
\centerline{\hspace{-3cm}\psfig{figure=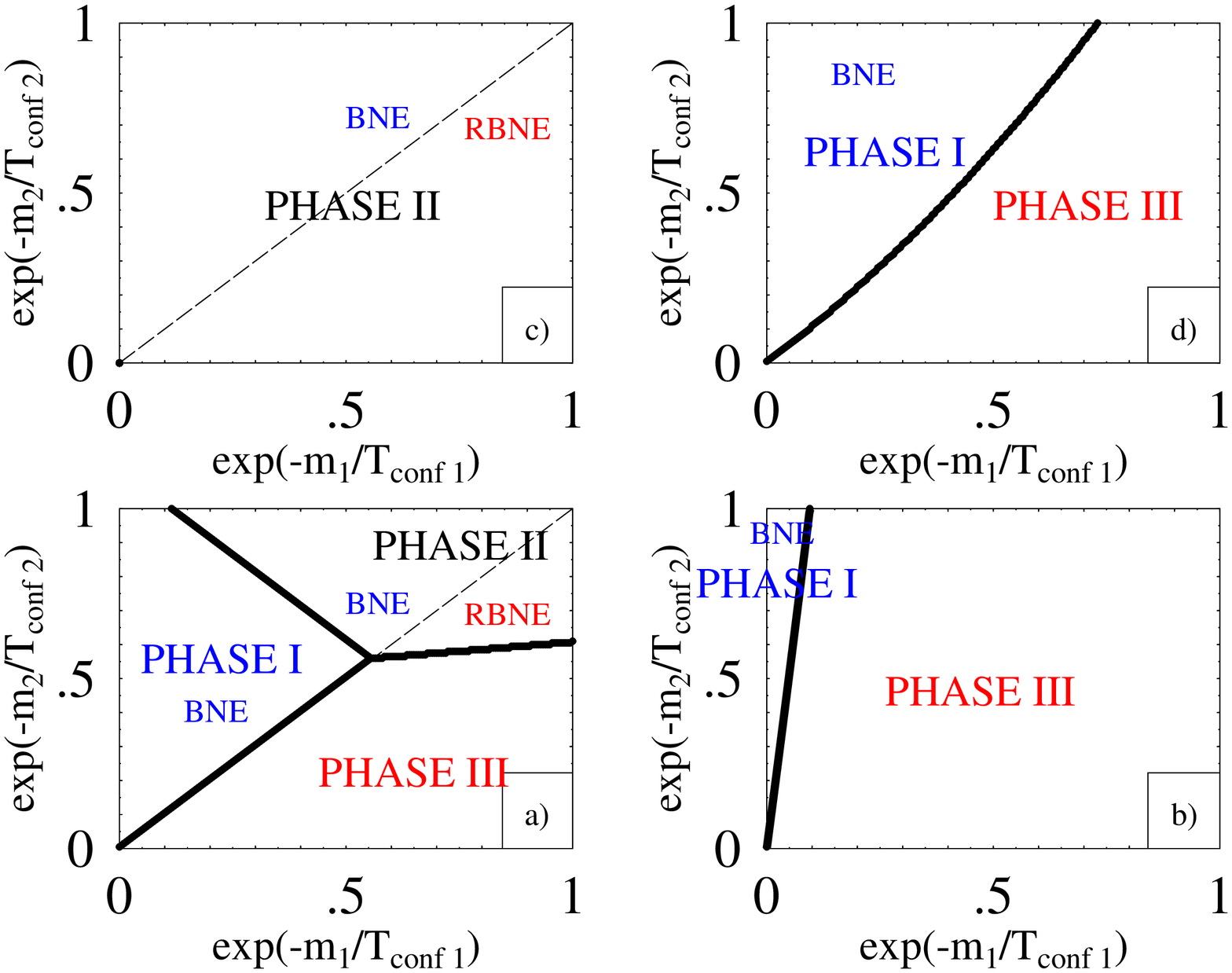,width=9cm,angle=-90}}
\vspace{-2cm}
\caption{
Phase diagrams of the mean field hard spheres mixture model 
in the plane $(\e^{-m_1\beta_1}, \e^{-m_2\beta_2})$ described in the text.
In panel $a)$, $b)$, $c)$, $d)$ the parameters
$(\e^{\mu_1},\e^{\mu_2})$ are respectively: $(.1,.1);(1,.1);(.1,1);(1,1)$.
In Phase I only BNE is present for any value of
$m_1\beta_1$ and $m_2\beta_2$. 
In Phase II crossover from BNE to RBNE occurs when
$m_1\beta_1/m_2\beta_2=1$ (the diagonal in the graph).
In Phase III large grains are expelled from the system.}
\label{fig_pd}
\end{figure}

\section{Phase Diagram and Segregation Phenomena}
\label{sect4}
After standard manipulation of eq.(\ref{zedw}), the saddle point 
free energy is obtained:
\begin{eqnarray}
f=-{\ln Z \over N} &=& -J%\left\{
\sum_{z=1}^{H}\left[%\left(
\sum_{j,k=1,2} x_A^{(j)}(z)%\right)\left(\sum_{k=1,2} 
x_B^{(k)}(z)%\right)
-x_{A}^{(1)}(z)x_{B}^{(1)}(z)\right]%\right\}
\nonumber \\&&
-\ln \left({\cal Z}[x_{A}]{\cal Z}[x_{B}]\right)
\label{f_gen}
\end{eqnarray}
where, after the $K\rightarrow\infty$ limit, we have defined
\begin{eqnarray}
{\cal Z}[x_{A}] &=& 1+\e^{-J[x_{A}^{(2)}(1)]}~\e^{\mu_1}
+\e^{-J[x_{A}^{(1)}(1)+x_{A}^{(2)}(1)]}~\e^{\mu_2}+
\e^{-J[x_{A}^{(2)}(1)+x_{A}^{(2)}(2)]}~\e^{2\mu_1-\beta_1m_1gh_0} \nonumber \\
&+&\e^{-J[x_{A}^{(1)}(1)+x_{A}^{(2)}(1)+x_{A}^{(2)}(2)]}~\e^{\mu_1+\mu_2-\beta_1m_1gh_0}+
\e^{-J[x_{A}^{(2)}(1)+x_{A}^{(1)}(2)+x_{A}^{(2)}(2)]}~\e^{\mu_1+\mu_2-\beta_2m_2gh_0} \nonumber \\
&+&\e^{-J[x_{A}^{(1)}(1)+x_{A}^{(2)}(1)+x_{A}^{(1)}(2)+x_{A}^{(2)}(2)]}~\e^{2\mu_2-\beta_2m_2gh_0}
\end{eqnarray}
Here we have introduced the density field $x_{a}^{(k)}(z)$, which is the 
number density fraction of species $k\in\{1,2\}$ on 
sublattice $a\in\{A,B\}$ at level $z\in\{1,...,H\}$. 
The constant $h_0$ is the vertical layer spacing. 
The $4H$ mean field self consistent coupled equations, 
to be solved in the limit $J\rightarrow\infty$, are the saddle points:
\begin{equation}
\frac{\partial f [x_A,x_B] } {\partial x_a^{(k)}(z)}=0
\label{sp}
\end{equation}
Without entering further details, eq.s(\ref{sp}) admit three kinds 
of solutions, corresponding to three different possible phases for the 
system. Phase I (which is twofold) is characterized by the property that 
there are no large grains sitting on the bottom layer, i.e., 
$x_{A}^{(2)}(1)=x_{B}^{(2)}(1)=0$. Phase II (twofold) admits the presence 
on both layers of both kind of particles.
Finally, Phase III has the property that small grains have expelled 
large grains, i.e., $x_{a}^{(2)}(z)=0$ $\forall a,z$. 
The resulting phase diagram of the system is shown in Fig.\ref{fig_pd}. 
In general the crossing from one to the other phase is marked by a 
first order phase transition. 
Finally, it is possible to show \cite{cdfnt} that the ``Bethe lattice'' 
version of the model used in the present ``fully connected'' mean field 
approach appears to be able to locate also a glassy phase in the above 
phase diagram, as originally predicted in \cite{NCH,cn_g,mimmo}. 

\begin{figure}[ht]
\centerline{\hspace{-1cm}\psfig{figure=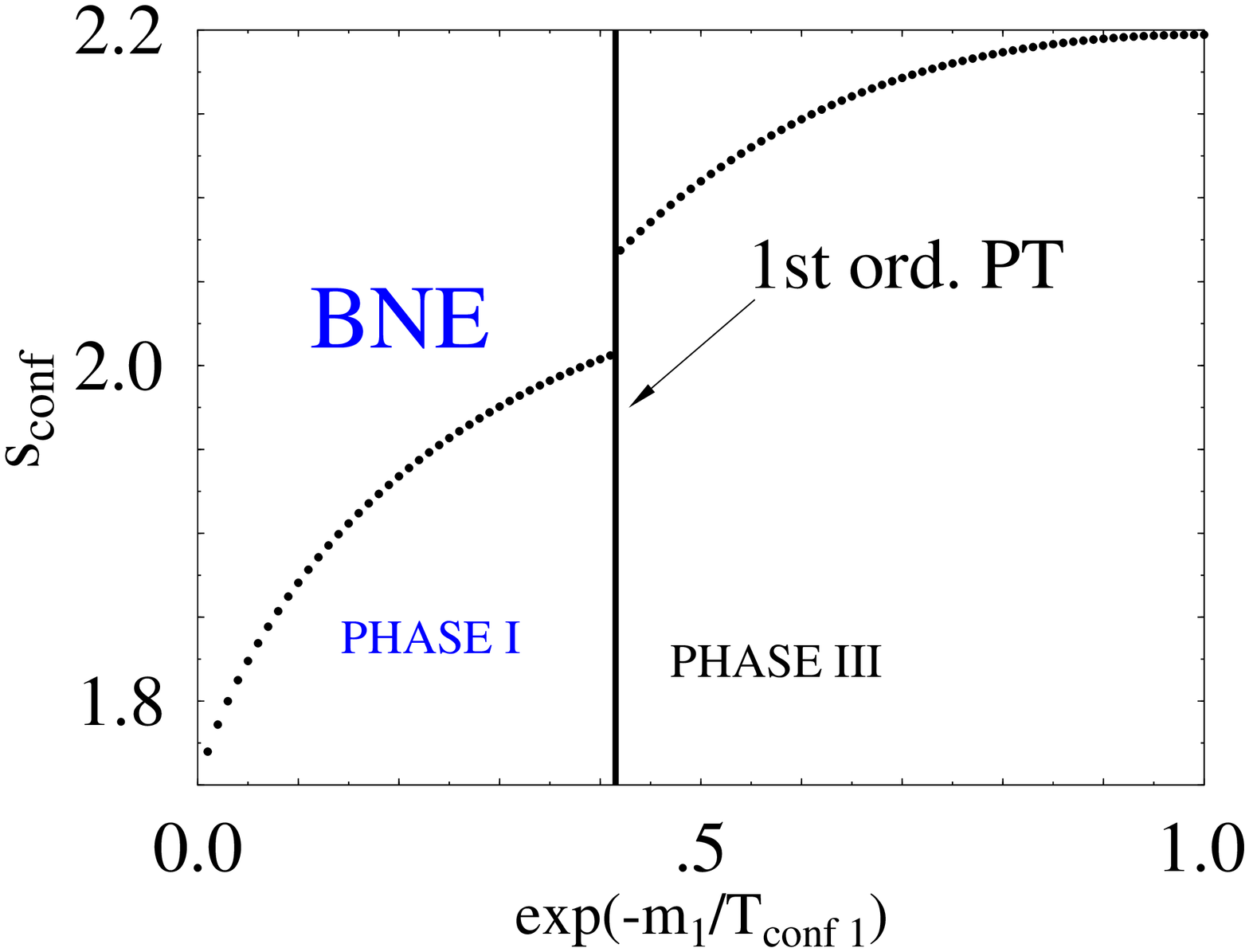,width=7cm,angle=-90}}
\vspace{-1.5cm}
\caption{The configurational entropy, $s_{conf}$, of the mean field model 
as a function of $\e^{-m_1\beta_1}$ for $(\e^{\mu_1},\e^{\mu_2})=(1,1)$ and 
$\e^{-m_2\beta_2}=.5$.}
\label{fig_s_c}
\end{figure} 

\begin{figure}[ht]
\centerline{
\hspace{-1cm}\psfig{figure=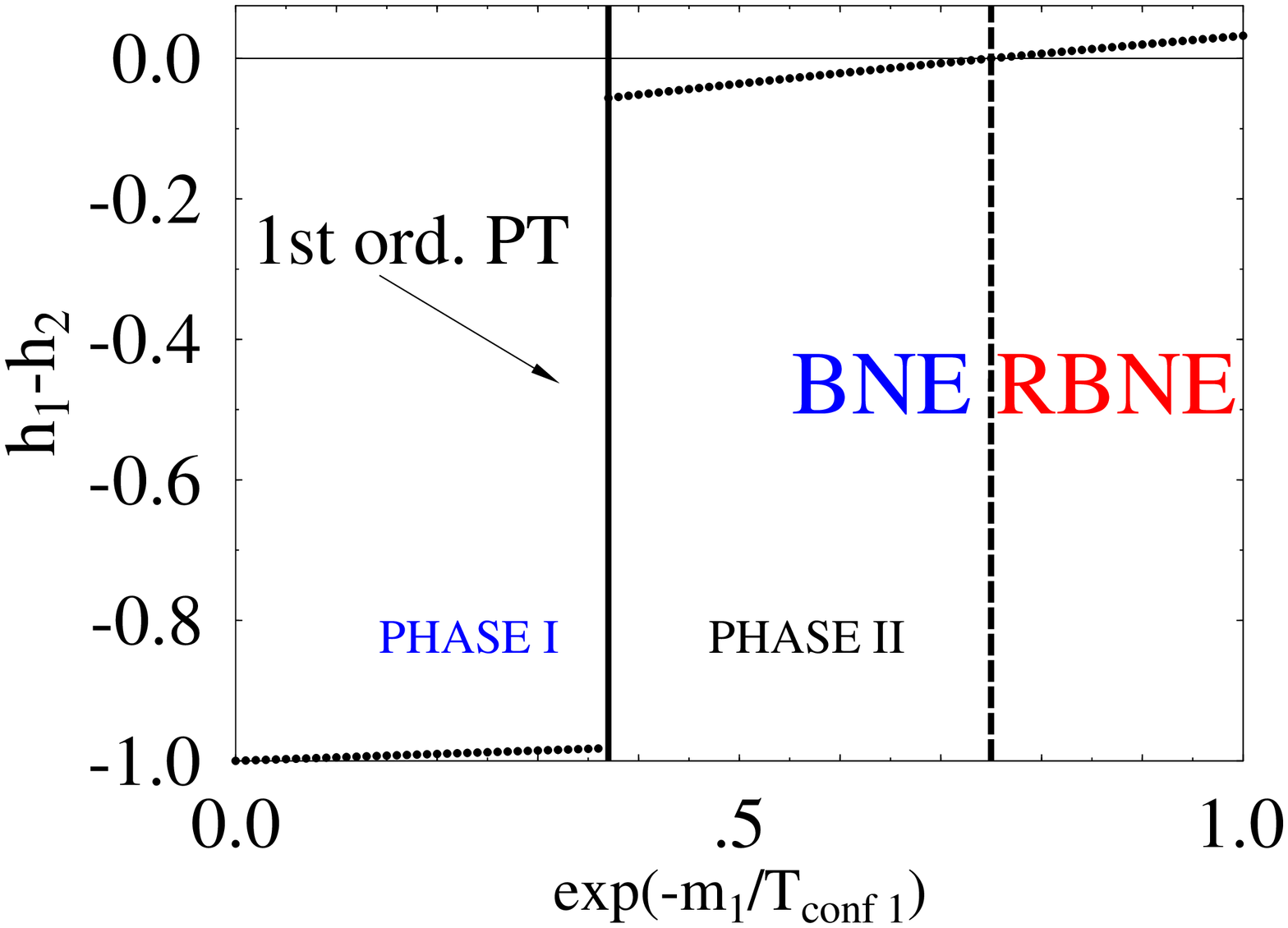,width=7cm,angle=-90}\hspace{1.5cm}
\psfig{figure=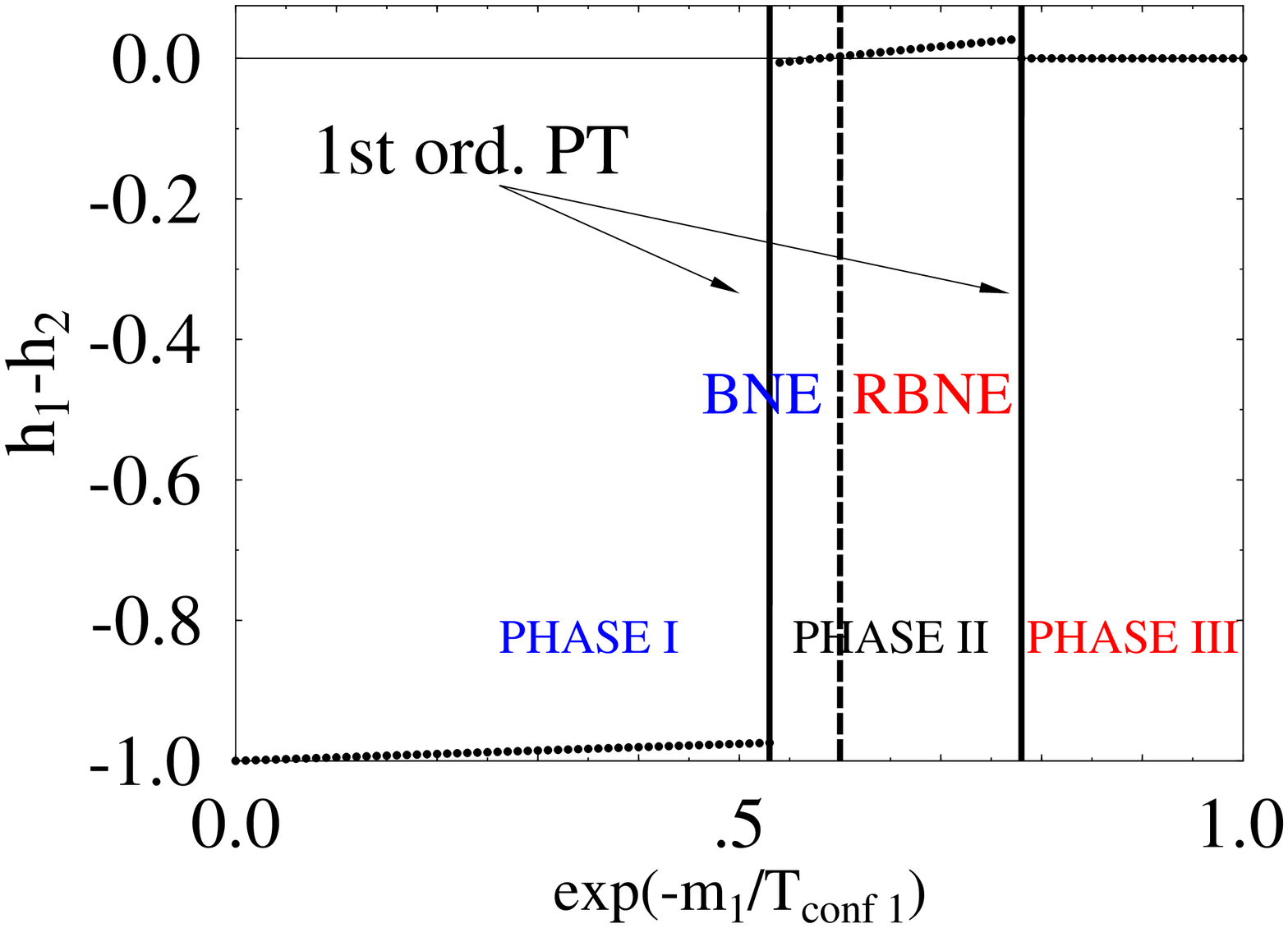,width=7cm,angle=-90}
}
\vspace{-1.5cm}
\caption{The vertical segregation parameter, $\Delta h=h_1-h_2$ (i.e., the 
average height difference of species 1 and 2) of the mean field model, 
as a function of $\e^{-m_1\beta_1}$ for $(\e^{\mu_1},\e^{\mu_2})=(.1,.1)$.
{\bf Left} panel: $\e^{-m_2\beta_2}=.75$;
{\bf right} panel: $\e^{-m_2\beta_2}=.58$.}
\label{fig_dh}
\end{figure}

It is worth to exhibit the configurational entropy of the system:
\begin{equation}
s_{conf}=\left[\beta_1\partial_{\beta_1}+\beta_2\partial_{\beta_2}+
\mu_1\partial_{\mu_1}+\mu_2\partial_{\mu_2}-1\right]f
\label{s_c}
\end{equation}
which is shown in Fig.\ref{fig_s_c}.
Finally, we discuss ``size segregation'', and specifically, 
the phenomenon known as the 
``Brazil Nut Effect'' (BNE), i.e., the tendency of large grains 
to segregate on top of the system. The vertical segregation parameter is the 
difference of average heights of species 1 and 2: $\Delta h=h_1-h_2$. 
It is possible to show that in Phase I large grains are always 
on average on top, i.e., $\Delta h<0$, due to the geometric organization of 
the pack. 
In Phase II, BNE is observed 
when $m_1\beta_1>m_2\beta_2$, else the ``Reverse Brazil Nut Effect'' (RBNE) 
is found (where large grains are on the bottom). 
Notice that in Phase II a route from the BNE to the RBNE region 
does not necessarily cut any transition line. Instead, crossing from the 
BNE region of Phase II 
to BNE region of Phase I (as in panel $a)$ of Fig.\ref{fig_pd}) corresponds 
to cross a 1st order phase transition line (with a jump in $\Delta h$, 
as shown in Fig.\ref{fig_dh}). Phase III is associated to an other form 
of segregation where, as already stated, small grains fully expel 
large grains. This phase separation phenomenon can give rise to ``vertical'' 
as well as ``horizontal segregation'', i.e., pattern formation.

Some numerical simulations \cite{fnc,rosato,luding} and experimental 
observations
\cite{bridgewater} about BNE appear to be explained by segregation mechanisms 
as those found in Phase I, II and III of the present model. 
In particular, segregation under ``condensation'', numerically discovered 
in \cite{luding}, may be originated by phenomena as those found in Phase III.

\section{Conclusions} 
As a theoretical picture is taking shape, there are still many open problems 
with the Statistical Mechanics of granular media; 
of course, the assumption that the stationary distribution is given by 
the maximum of an entropy with a few macroscopic constraints is to 
be checked case by case; and there is no general a priori criteria 
to select such necessary macroscopic constraints, i.e., there is no general 
criteria to establish how many ``thermodynamic'' parameters ($T_{conf}$'s) 
are needed to characterize the status of a given powder \cite{fnc,berg}. 

Anyway, simple models of granular media have confirmed and extended 
\cite{fnc,mimmo,kurchan,Dean,brey} Edwards' 1989 hypothesis \cite{Edwards} 
that granular media in their ``inherent states'' can be described by a 
generalized Statistical Mechanics. 
Here we have reviewed recent progresses made in this research area and, 
in particular, we have analytically investigated a simple mean field model 
where to apply Edwards' Statistical Mechanics approach. 
A comprehensive picture emerges of ``thermodynamical'' and dynamical 
properties of these non thermal systems, ranging from segregation patterns 
to their jamming transition \cite{Edwards,fnc,cn}. The presence of phase 
transitions in the inherent states space, here predicted 
by a mean field theory, is to be confirmed by experimental work.


\begin{thebibliography}{99}
\bibitem{note_p} When more than two layers are considered the projector 
term, for $z>2$, becomes:
\begin{equation}
{\cal H}_{Edw}(z)= K \sum_{i,a}\left[
(1-\delta_{n_i^{a_z},0})\delta_{n_i^{a_{z-1}},0}\delta_{n_i^{a_{z-2}},0}+
\delta_{n_i^{a_z},1}\delta_{n_i^{a_{z-1}},0}\delta_{n_i^{a_{z-2}},1}
\right]
\label{H_pi2}
\end{equation}
Grains on the bottom layer ($z=1$) are always considered to be stable 
(i.e., ${\cal H}_{Edw}(z=1)=0$).
\end{thebibliography}
\end{document}